\documentclass[%
 aip,apl,
 amsmath,amssymb,
 reprint,%
]{revtex4-2}

\usepackage{graphicx}
\usepackage{dcolumn}
\usepackage{bm}

\usepackage[utf8]{inputenc}
\usepackage[T1]{fontenc}
\usepackage{mathptmx}

\usepackage{upgreek}

\begin{document}


\title{Domain wall motion at low current density in a synthetic antiferromagnet nanowire}

\author{Christopher~E.~A.~Barker}
\affiliation{School of Physics and Astronomy, University of Leeds, Leeds LS2 9JT, United Kingdom}

\author{Simone~Finizio}
\affiliation{Paul Scherrer Institut, 5232 Villigen PSI, Switzerland}

\author{Eloi~Haltz}
\altaffiliation{Present address: Universit\'{e} Sorbonne Paris Nord, 99 Av. Jean Baptiste Cl\'{e}ment, 93430 Villetaneuse, France}
\affiliation{School of Physics and Astronomy, University of Leeds, Leeds LS2 9JT, United Kingdom}

\author{Sina~Mayr}
\affiliation{Paul Scherrer Institut, 5232 Villigen PSI, Switzerland}
\affiliation{Laboratory for Mesoscopic Systems, Department of Materials, ETH Zurich, 8093 Zurich, Switzerland}

\author{Philippa~M.~Shepley}
\affiliation{School of Physics and Astronomy, University of Leeds, Leeds LS2 9JT, United Kingdom}
\affiliation{Bragg Centre for Materials Research, University of Leeds, Leeds LS2 9JT, United Kingdom}

\author{Thomas~A.~Moore}
\affiliation{School of Physics and Astronomy, University of Leeds, Leeds LS2 9JT, United Kingdom}

\author{Gavin~Burnell}
\affiliation{School of Physics and Astronomy, University of Leeds, Leeds LS2 9JT, United Kingdom}

\author{J\"{o}rg~Raabe}
\affiliation{Paul Scherrer Institut, 5232 Villigen PSI, Switzerland}

\author{Christopher~H.~Marrows}
\email{c.h.marrows@leeds.ac.uk}
\affiliation{School of Physics and Astronomy, University of Leeds, Leeds LS2 9JT, United Kingdom}
\affiliation{Bragg Centre for Materials Research, University of Leeds, Leeds LS2 9JT, United Kingdom}

\date{\today}

\begin{abstract}
The current-driven motion of magnetic domain walls (DWs) is the working principle of magnetic racetrack memories. In this type of spintronic technology, high current densities are used to propel DW motion in magnetic nanowires, causing significant wire heating. Synthetic antiferromagnets are known to show very fast DW motion at high current densities, but lower current densities around onset of motion have received less attention. Here we use scanning transmission x-ray microscopy to study the response of DWs in a SAF multilayer to currents. We observe that the DWs depin at $\sim 3 \times 10^{11}$~A/m$^2$ and move more quickly in response to 5~ns duration current pulses than in comparable conventional multilayers. The results suggest that DWs in SAF structures are superior to conventional N\'{e}el DWs for low energy consumption racetrack technologies.
\end{abstract}

\maketitle

Magnetic domain walls (DWs) separate uniformly magnetized domains in a ferromagnet. They are narrow regions where the magnetization rotates between the directions in the domains and both influence and respond to spin-polarised currents \cite{Marrows2005}. The use of a spin-polarized electron flow and its resulting torques on magnetic textures has been demonstrated for driving DWs in magnetic wires. New generations of  devices have been proposed based on these effects, such as the so-called magnetic racetracks that can be used as storage-class memories \cite{Parkin2008,Parkin2015} or for novel forms of information processing \cite{Luo2020,Vakili2020,Ollivier2021}. In this type of technology, a stream of bits is encoded as a series of domains separated by domain walls in a magnetic nanowire, which can be shifted along the wire using electrical current pulses by means of the spin-torque mechanism. The DW velocity influences the speed of operation whilst the power dissipated by the current pulse influences the energy consumption of the device.

Initial versions of the magnetic racetrack were developed using a simple strip of an in-plane magnetised soft magnetic material such as Permalloy in which the domain wall motion was actuated by the volume spin-transfer-torque \cite{Parkin2008}. Further generations have incorporated developments such as multilayer wires with interface-induced effects such as perpendicular magnetic anisotropy (PMA) and   Dzyaloshinskii-Moriya interactions (DMI) to enforce wall chirality \cite{Parkin2015}, with domain wall motion now being driven by spin-orbit torques \cite{Emori2013,Ryu2013}.

Most recently, synthetic antiferromagnets (SAFs) \cite{Duine2018} have been introduced as race track materials \cite{Yang2017}. These systems are composed of two ferromagnetic layers coupled to each other antiferromagnetically through a non-magnetic spacer layer. The magnetic moments of the two layers compensates for each other leading to fast magnetization dynamics. In particular, very fast domain wall motion at several 100~m/s for high current drives of a few~TA/m$^2$ have been observed in these systems using a Ru spacer \cite{Yang2015}. It has subsequently been shown that replacing the spacer material with Rh reduces the DW velocity \cite{Cohen2020}, and that DW velocity in a SAF can be controlled by means of iontronic gating \cite{Guan2021}.


Power consumption is a key constraint for such applications. In particular, fast motion at low current densities, just above the critical current density $J_\mathrm{c}$ at which DWs depin and can be set in motion, is critical. Here we study, using scanning transmission x-ray microscopy (STXM), the low current density dynamics of DWs around the onset of motion. We show that $J_\mathrm{c} \sim 3 \times 10^{11}$~A/m$^2$ and that DWs move more quickly at lower current densities than in comparable conventional multilayers. The multilayers that we study contain ten magnetic layers, showing that the advantages of the SAF structure extend beyond a pair of magnetic layers separated by a single spacer layer.

The SAF multilayers that we studied were deposited by magnetron sputtering in a chamber with a base pressure of $1.0 \times 10^{-9}$~mbar under a working pressure of $4.5 \times 10^{-3}$~mBar of Ar. Typical deposition rates, calibrated by x-ray reflectometry on test films, were 0.5~\AA /s. Multilayers were grown simultaneously on x-ray transparent Si$_3$N$_4$ membranes for transmission microscopy and solid thermally oxidised Si substrates for magnetometry measurements.

\begin{figure}[t]
\includegraphics[width=7.5cm]{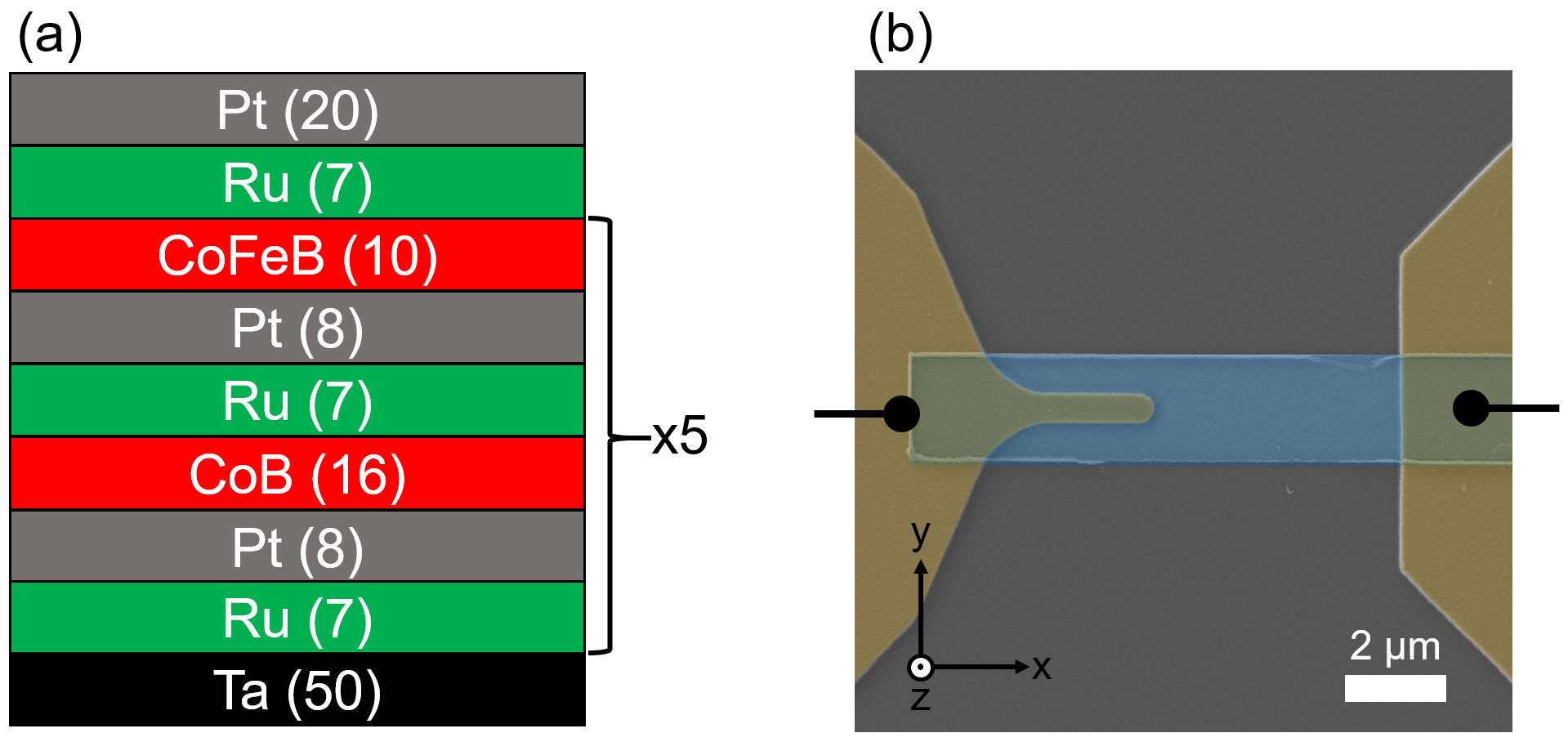}
\caption{\label{fig:stack} (a) SAF multiayer stack structure, with layer thicknesses given in \AA; (b) SEM image of device.}
\end{figure}

The SAF multilayer stack that we studied is shown in Fig.~\ref{fig:stack}(a). It features alternating layers of Co$_{68}$B$_{32}$ and Co$_{40}$Fe$_{40}$B$_{20}$, which form oppositely magnetised sublattices in the SAF ground state, owing to the indirect antiferromagnetic exchange through the Ru/Pt spacers, whose thicknesses were chosen to ensure this form of coupling\cite{Lavrijsen2012}. There are five pairs of such layers in the completed stack. The ferromagnetic (FM) layer thicknesses were selected so that they have equal and opposite magnetic moments. PMA and DMI are induced in the layers at their interfaces, predominantly by the heavy metal Pt.

Multilayers on the membranes were patterned into 2~$\upmu$m wide wires by electron beam lithography and lift off, with current contacts at either end made from Cu, which is a light enough element to remain transparent to x-rays at the Co and Fe $L_3$ absorption edges. The design is similar to that used for skyrmion injection in Ref. \onlinecite{Finizio2019}, with a wide drain contact at the right hand end and a finger-shaped source contact on the left.

\begin{figure}
\includegraphics[width=8cm]{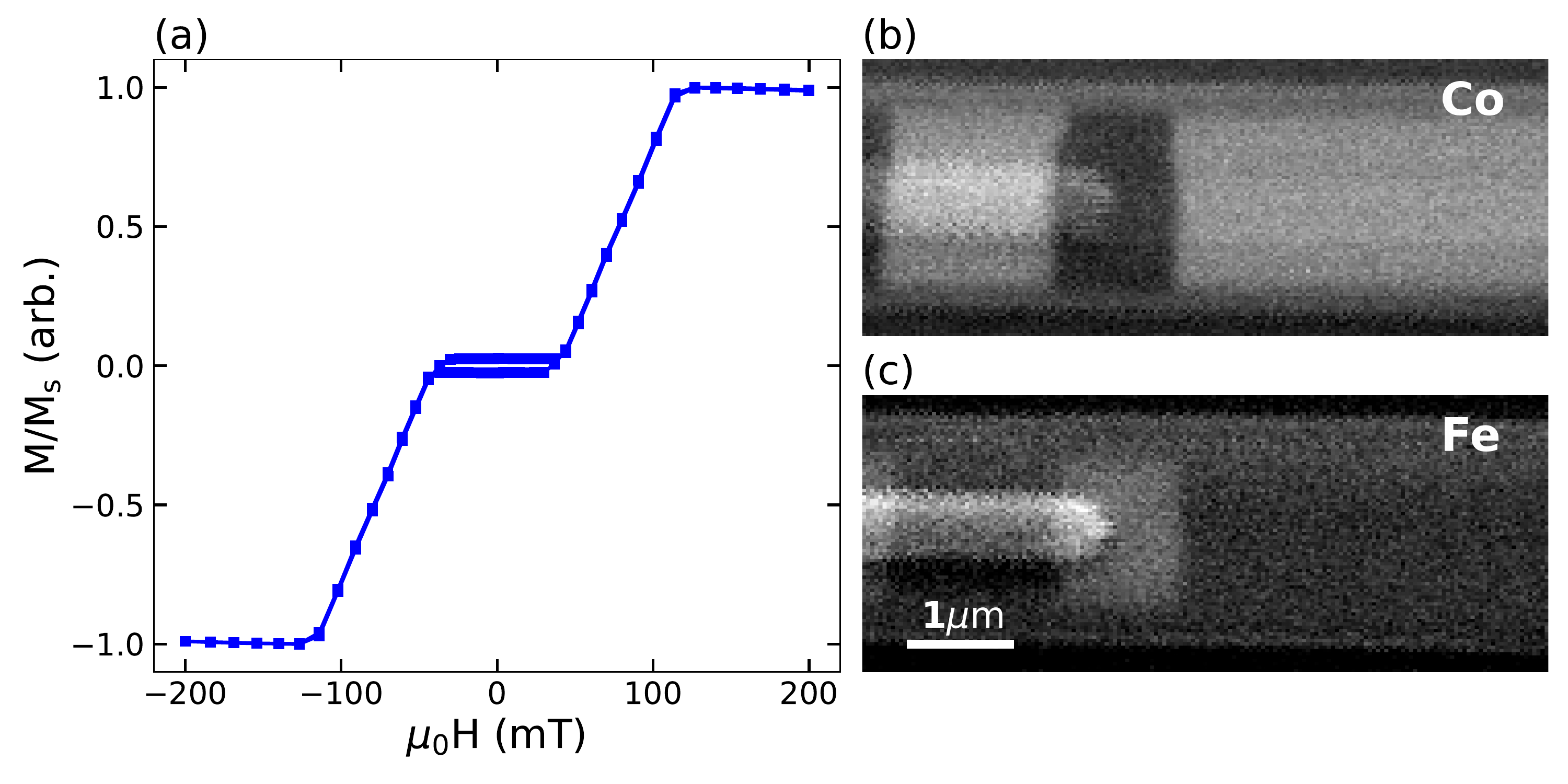}
\caption{\label{fig:squid} (a) SQUID-VSM hysteresis loop; (b),(c) STXM images showing opposite contrast at the Co and Fe $L_3$ edges, respectively.}
\end{figure}

The multilayer on the solid substrate was characterized magnetically using superconducting quantum interference device vibrating sample magnetometry (SQUID-VSM). A magnetic hysteresis loop, acquired at room temperature and normalized to the saturation magnetization, is shown in fig.~\ref{fig:squid}. For applied fields smaller than about 45~mT the magnetisation is constant and very small, less than 10~\% of the saturated value, showing that stack does indeed have a SAF ground state, and that the relative FM layer thicknesses are indeed close to the balance point.

The samples on membranes were imaged by means of STXM at the PolLux beamline at the Swiss Light Source. Images  of the region of the microwire close to the source contact are shown Fig.~\ref{fig:squid}(b) and (c), acquired using x-ray photons with energies at the Co and Fe $L_3$ absorption edges, respectively. Since magnetic contrast in this technique arises owing to the x-ray magnetic circular dichroism effect (XMCD), we can separately probe the magnetism of these two elements by this means. The x-ray beam passed through the multilayer at normal incidence, and so magnetic contrast arises that is proportional to the out-of-plane component of the magnetization, $M_z$.

In each image a domain that spans the width of the wire can be observed under the tip of the source contact, appearing as a band of dark contrast in Fig.~\ref{fig:squid}(b) and a band of light contrast in Fig.~\ref{fig:squid}(c). In this case we are imaging the as-patterned state, prior to the application of any fields or current pulses. Since only one sublattice of FM layers, made from CoFeB, contains Fe, only those layers contribute to the contrast in Fig.~\ref{fig:squid}(c). Meanwhile, the Co content of the CoB layers is higher than that of the CoFeB layers, so that although the contrast in Fig.~\ref{fig:squid}(b) arises from both sublattices, it will be predominantly from the CoB sublattice. The fact that the contrast is of opposite sign in this pair of images confirm the SAF state, with CoB and CoFeB sublattices being oppositely magnetized.

\begin{figure}
\includegraphics[width=8.5cm]{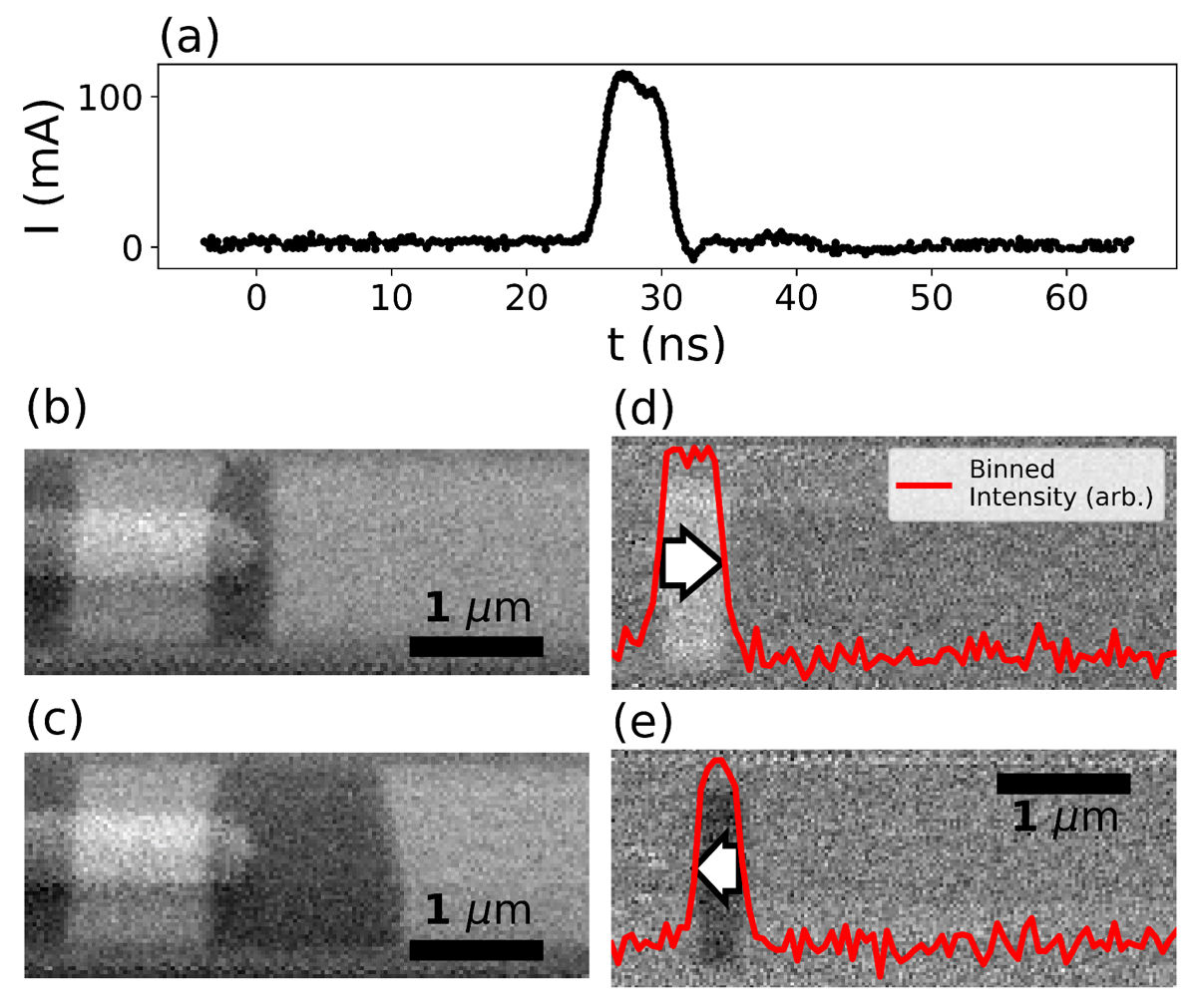}
\caption{\label{fig:dwmotion} (a)pulse from oscilloscope. (b), (c) STXM images showing before and after images of DW motion, (d),(e) difference images showing DW motion and fitting.}
\end{figure}

We have studied the response of such DWs to nominally 5~ns duration pulses of electrical current injected into the magnetic wire.  The pulses were generated with a Keysight M8195A arbitrary waveform generator (64~GSa/s sampling rate) combined with a SHF-826H amplifier. An oscilloscope (50~$\Omega$-terminated) trace of such a pulse after it has passed through the wire is shown in Fig.~\ref{fig:dwmotion}(a). The pulse retains a high degree of squareness with modest ringing after the pulse, which peaks at a current density $J \sim 3.0 \times 10^{11}$~A/m$^2$.

Fig.~\ref{fig:dwmotion}(b) and (c) show STXM images acquired before and after the application of five such pulses, separated by a few seconds to allow for a complete cooling of the magnetic wire. The positions of the DWs are marked with dashed red lines. The left-hand DW is unaffected by the current pulses, as expected since the current is shunted by the overlying Cu finger electrode. On the other hand, the right hand wall moves away from the finger electrode, in the direction of the conventional current flow, indicating spin-orbit torques as the most likely mechanism to drive the motion. The movement is a distance of a few hundred nm, and is not completely uniform across the wire, with the DW moving further along the edge of the wire in the lower part of the image than the edge in the upper part in this case.

This sort of DW tilting under spin-orbit torques has been previously observed in Co/Ni/Co multilayers with DMI \cite{Ryu2012}, subsequently shown to be an intrinsic effect of the dynamics of DWs under DMI \cite{Boulle2013}. The model of Ref.~\onlinecite{Boulle2013} predicts that the tilt angle should be $\propto M_\mathrm{s}$ and so ought to be zero for a SAF. Whilst it is quite common for us to see some tilt in our images like the example in Fig.~\ref{fig:dwmotion}(c), the tilt angle is quite variable and can be of either sign. Based on these considerations, we think it is likely that the tilts we observe are extrinsic and arise stochastically from pinning of the DWs at defects, particularly on lithographic imperfections at the edges of our microwires.

In order to study this DW motion in a more systematic way, we acquired a series of images using positive helicity photons at the Co $L_3$ edge before and after series of current pulses of different $J$ but with the same duration, 5~ns. Subtracting pairs of these images yields difference images: examples are shown in Fig.~\ref{fig:dwmotion}(d) and (e). These difference images show the area swept out by the DW during motion driven by sequences of 10 positive polarity (d), and then 10 negative polarity (e) pulses of $6.35 \times 10^{11}$~A/m$^2$. Bright contrast in the difference corresponds to rightward motion, and dark contrast to leftward motion.

To determine the distance traveled by the DW, the contrast values in a column of pixels across the wire were binned together. The red lines in Fig.~\ref{fig:dwmotion}(c) and (d) show how this binned contrast varies along the length of the wire, with clear excursions away from the background level in the region where the DW motion has occurred. The distance traveled by the DW is then taken to be the full-width at half maximum value of this excursion, giving the average value for distance traveled in the case of non-uniform wall motion. We determined the velocity as this distance divided by the total nominal duration of the pulses in the pulse sequence, and so use this term to refer to the average velocity of the domain wall during its motion.

\begin{figure}
\includegraphics[width=8cm]{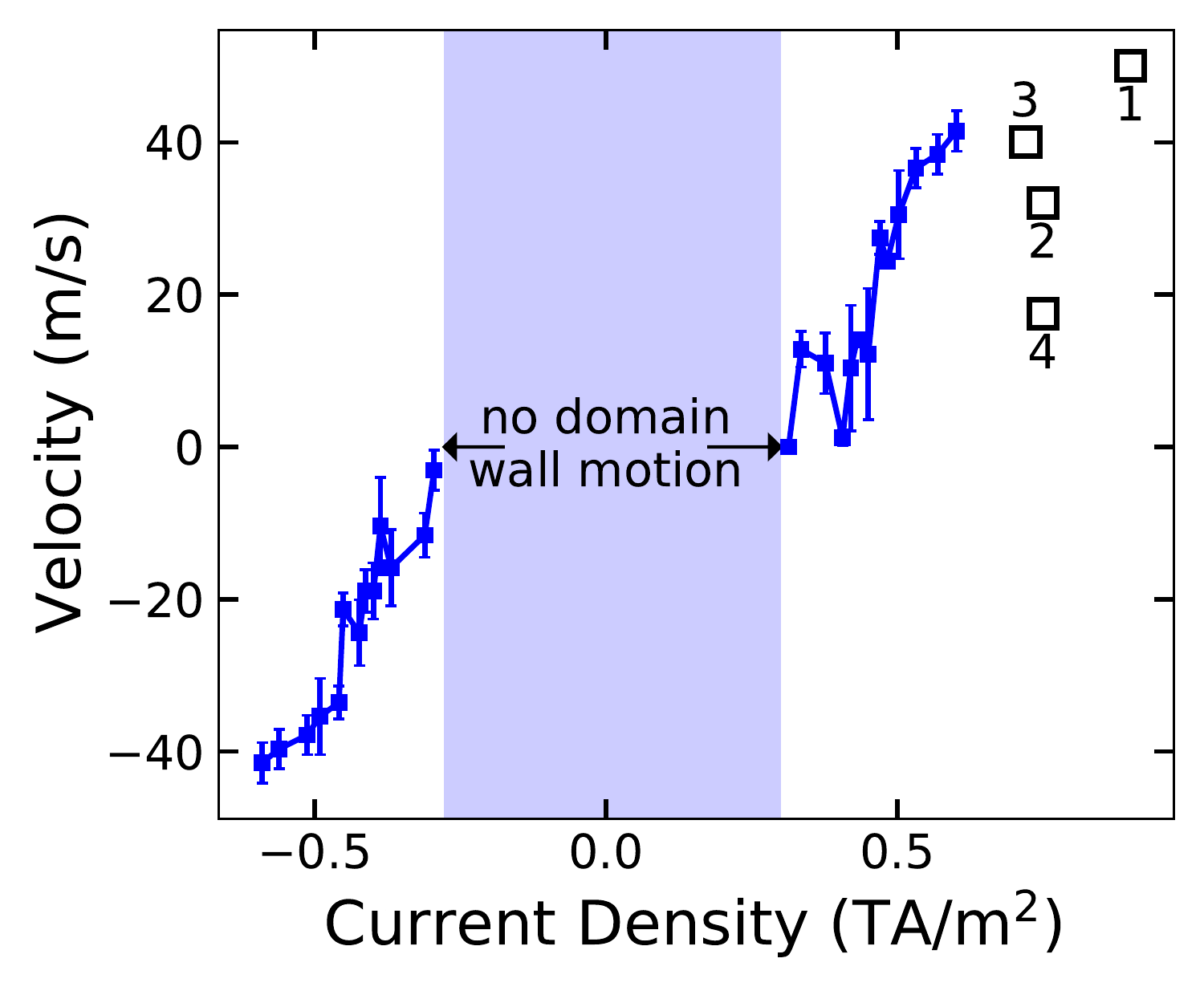}
\caption{\label{fig:dwvelocity} DW velocity vs current density. Red data points are derived from prior work on comparable CoB-based samples: data point [1] is derived from data reported in Ref.~\onlinecite{Finizio2018}, data points [2] and [4] from Ref.~\onlinecite{Finizio2020a}, and data point [3] from Ref.~\onlinecite{Finizio2020}.}
\end{figure}

The velocities derived from this analysis are plotted as a function of current density in Fig.~\ref{fig:dwvelocity}. There is a region of low $J$ where there is no observable DW motion. Once $|J|$ exceeds a critical value of $J_\mathrm{c} \sim 3 \times 10^{11}$~A/m$^2$ then DW motion in the direction of conventional current flow takes place. We only measured up to a values of $J = 6 \times 10^{11}$~A/m$^2$ owing to concerns about the thermal stability of our nanostructure on the poorly heat-sunk membrane substrate. Nevertheless, the DW velocity rises as the current becomes stronger, and reaches a value just exceeding 40~m/s at that maximum current density.

To compare the results we have obtained here on DWs in SAF multilayers we have performed a meta-analysis on prior work on STXM imaging of current-driven DW motion in conventional multilayers based around a Pt/CoB/Ir repeat unit that have a FM-aligned ground state \cite{Finizio2018,Finizio2020,Finizio2020a}. Data points obtained from those previous works are also plotted in Fig.~\ref{fig:dwvelocity}.  We can see that comparable velocities were reached in those experiments, but substantially higher current densities were required to do so. Our results mirror a reduction in $J_\mathrm{c}$ seen previously in in-plane magnetized SAFs when compared to simple Permalloy wires \cite{Lepadatu2017}.

The depinning current observed here is also about half of that reported by Yang et al. in Ref. \onlinecite{Yang2015} for a two-layer SAF. In that case, only one of the two layers had an interface with a heavy metal--Pt--to supply both DMI and spin-orbit torque. An important difference is that we have more magnetic layers in our SAF and every one of them is in contact with Pt to supply these two critical effects.

To summarise, we have studied DW motion in a SAF multilayer using STXM. We find motion at comparable velocities to conventional perpendicularly magnetised multilayers, but at reduced current densities that imply power dissipation reduced by up to 50 \%. We also find a critical current density $J_\mathrm{c}$ that is substantially less than that in a two-layer SAF, indicating the advantages of a stack design where every magnetic layer is subject to a spin-orbit torque. Our results offer promise for low power consumption DW racetrack technologies as well as suggesting that the motion of skyrmions in such SAFs \cite{Dohi2019,Legrand2020,Chen2020} is also likely to only require modest current densities.

\begin{acknowledgments}
This work was supported by the UK EPSRC (grant numbers EP/T006803/1 and EP/R00661X/1). We acknowledge support from the Sir Henry Royce Institute for access to the the Royce Deposition System for sample growth. C.E.A.B. acknowledges support from the National Physical Laboratory. Part of this work was performed at the PolLux (X07DA) beamline of the Swiss Light Source, Paul Scherrer Institut, Villigen PSI, Switzerland. The research leading to these results received funding from the European Community's Seventh Framework Programme (No. FP7/2007-2013) under Grant Agreement No. 290605 (PSI-FELLOW/COFUND), the Swiss National Science Foundation under Grant Agreement No. 172517.
\end{acknowledgments}

\section*{Author Declarations}
\subsection*{Conflict of Interest}
The authors have no conflicts to declare.

\section*{Data Availability}
Data pertaining to this work is available from the Research Data Leeds repository at https://doi.org/TBD.

\section*{References}


%

\end{document}